\RequirePackage{fix-cm}
\documentclass{svjour3}                     
\smartqed  
\usepackage{graphicx}
\newcommand{\braket}[1]{ \langle #1 \rangle}

\newcommand{\bv}[1]{\mbox{\boldmath $#1$}}
\begin{document}

\title{Two-neutron halo structure and anti-halo effect in $^{31}$F 
}


\author{Hiroshi Masui \and Wataru Horiuchi  \and Masaaki Kimura 
}


\institute{H. Masui \at
              Information Processing Center, Kitami Institute of
              Technology, Kitami, 090-8507, Japan \\
              \email{hgmasui@mail.kitami-it.ac.jp}           
           \and
             W. Horiuchi \at
             Department of Physics, Hokkaido University, Sapporo
             060-0810, Japan
           \and
             M. Kimura \at
             Department of Physics, Hokkaido University, Sapporo
             060-0810, Japan and \\
             Nuclear Reaction Data Centre, Hokkaido University, Sapporo
             060-0810, Japan and \\
             Research Center for Nuclear Physics (RCNP), Osaka
             University, Ibaraki, 567-0047, Japan 
}

\date{Received: date / Accepted: date}

\maketitle

\begin{abstract}
We perform a detailed analysis for the structure of $^{31}$F,
which is a candidate of the halo nucleus.
We calculate the radius and reaction cross-section using
a three-body model of $^{29}$F+$n$+$n$
and discuss how the competition between the neutron-pairing
and the single-particle energy induces structural changes of $^{31}$F.
The present analysis further clarifies a new aspect of
the anti-halo effect that suppresses the halo structure.
\end{abstract}

\section{Introduction}
\label{intro}
Characteristic nuclear structure with loosely bound nucleons
is called ``halo'' and has been extensively studied from both the theoretical
and experimental sides~\cite{Ta13_PPNP,Zh93_PR231}.
In recent years, 
a drip-line nucleus of the fluorine isotopes $^{31}$F
has been observed~\cite{Ah19_PRL123}.
Because of the small neutron separation energy
and the subsystem $^{30}$F ($^{29}$F+$n$) being unbound,
$^{31}$F is expected to have a halo structure 
same as other Borromean nuclei,
e.g. $^{6}$He, $^{11}$Li, $^{14}$Be, and $^{19}$B.
On the other hand,
although  development of the halo structure is closely related
to the small neutron separation energy,
a mechanism to suppress the nuclear radius in weakly bound systems
has been discussed as the ``pairing anti-halo'' effect~\cite{Be00_PLB496,Ha11_PRC84}.

Therefore, in order to give a theoretical prediction to the structure of $^{31}$F,
we performed three-body calculations using a $^{29}$F+$n$+$n$  model
and showed the matter radius and reaction cross-sections of $^{31}$F
in Ref.~\cite{Ma20_PRC101}.
This analysis revealed that 
the gap between the $p$-orbit and $f$-orbit in the subsystem $^{30}$F
is the essential ingredient to determine the radius of $^{31}$F.
Furthermore, a new aspect to the anti-halo effect was proposed.

In this paper, we discuss the mechanism of the newly proposed
anti-halo effect of $^{31}$F more detail
and clarify a general condition for the formation of the halo
structure and the occurrence of the anti-halo effect.

\section{Theoretical model and interactions}
\label{sec:1}
We calculate the structure of $^{31}$F in a $^{29}$F+$n$+$n$
three-body model.
To describe $^{31}$F as a loosely-bound Borromean system, 
the calculation method is required to be able to
accurately take into account the continuum states.
For this purpose,
we employ the cluster-orbital shell model (COSM)~\cite{Su88_PRC38}
with the Gaussian expansion method (GEM)~\cite{My14_PPNP79,Ma14_PRC89}.

The Hamiltonian and basis functions in COSM with GEM
for the $^{29}$F+$n$+$n$ system are briefly explained below.
The  Hamiltonian by removing the center of mass motion
in COSM is formulated as follows:
\begin{equation}
  \label{eq:COMS_Hamiltonian}
 \hat{H} = \sum_{i=1}^{2}
  (\hat{T}_{i} + \hat{V}_{i} + \lambda \hat{\Lambda}_{i})
  +(\hat{t}_{12} + \hat{v}_{12}) .
\end{equation}
Here, $\lambda \hat{\Lambda}$ is introduced to eliminate
the Pauli forbidden states in the procedure of
the orthogonality condition model~\cite{Sa77_PTPS62},
and $\hat{t}_{12}$ is the recoil term induced
by the subtraction of the center of mass motion.

We solve the eigenvectors of the Hamiltonian (\ref{eq:COMS_Hamiltonian})
using the Gaussian basis $\Phi_{pq\ell j}^{JM}$ in a variational way as 
$ \Psi_{JM} = \sum_{pq\ell j}  C_{pq\ell j} \Phi_{pq\ell j}^{JM}$.
Since we assume the core nucleus $^{29}$F as a spin-less particle for simplicity,
the spin-parity of $^{31}$F is determined by the valence nucleons part
$\Phi_{pq\ell j}^{JM}    \equiv   {\mathcal A}
  \left\{ [
  \phi_{p\ell j}(1) \otimes\phi_{q\ell j}(2)
            ]_{JM}
\right\} $.
For the basis functions, we take the maximum orbital angular momentum
up to 5, and the width parameters of the Gaussian are prepared using
a geometric progression manner,
where 20 bases are applied for each coordinate.
Therefore, for example,
a typical basis size is 2310 for the $0^{+}$ state.

For the $^{29}$F+$n$ interaction $\hat{V}_{i}$,
we employ the Woods-Saxon (WS) potential using the same parameter sets
for reproducing the reaction cross-sections of
$^{31}$Ne~\cite{Ho10_PRC81},
which is the next nucleus of $^{30}$F by a proton.
Since there is no information other than $^{30}$F being unbound,
we adjust the strength parameter $V_{0}$ using two different values
for each set of $\{ r_{0} , a \}$ in WS
under the condition that $^{30}$F is unbound and $^{31}$Ne is bound.
This leads to 12 different parameter sets for the calculation~\cite{Ma20_PRC101}.
For the neutron-neutron potential $\hat{v}_{12}$, we use the central
part of the Minnesota potential~\cite{Th77_NPA286}
with the exchange parameter $u=1.0$.

\section{Results and Discussion}
\label{sec:3}
First, we perform calculations using the 12 different parameter sets.
As a result, the two-neutron separation energies $S_{2n}$ are obtained 
in the range of $0.44$ -- $1.37$ MeV, and the matter radius $R_{\rm rms}$ 
varies in the range of $3.48$ -- $3.70$ fm.
For the maximum radius case $R_{\rm rms} = 3.70$ fm,
the halo structure is considered to be well developed.
We also calculate the reaction cross-sections
at the incident energies $240$ and $900$ MeV/nucleon
using the Glauber model (NTG)~\cite{Glauber,NTG,Abu08_PRC77,Ho12_PRC86}
and obtain large cross-sections as $1530$ and $1640$ mb~\cite{Ma20_PRC101}.
This result further supports the large matter radius of $^{31}$F,
i.e. the halo structure,
is realized under a certain condition in $^{30}$F.

\begin{figure*}
  \includegraphics[width=0.73\textwidth]{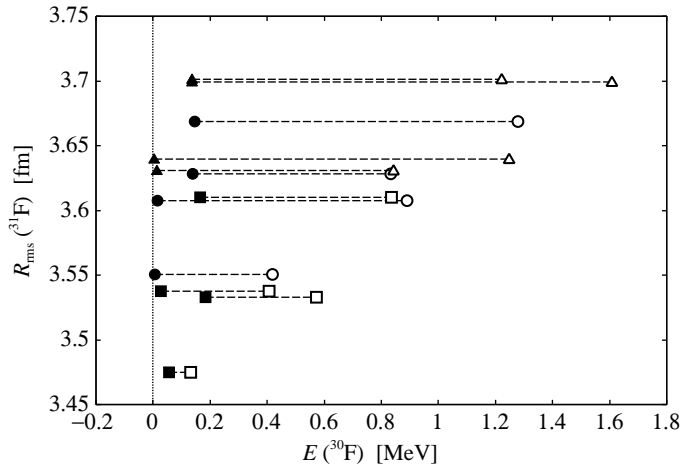}
\caption{Calculated matter radius of $^{31}$F in different parameter
  sets of $\{ V_{0}, a, r_{0} \}$ with the WS potential of $^{30}$F.
  The $x$-axis shows the energy of $^{30}$F, and the triangles, circles
  and squares are the results for the parameter sets
  of $r_{0} = 0.65$, $0.70$, and $0.75$ fm, respectively.
  Dashed-line indicates the pair of $^{30}$F energies for $p_{3/2}$
  (solid) and $f_{7/2}$ (open) orbits.
  For detail, see text.}
\label{fig:E30F_Rrms}    
\end{figure*}

Next, to clarify the condition,
we analyze the characteristics of the results
for $^{31}$F using the 12 parameter sets.
To this end, we introduce the Pearson's correlation coefficient (PCC)
between two variables $x$ and $y$, 
\begin{equation}
  \label{eq:PCC}
    r_{xy} = \frac{
   \sum_{i = 1}^{M} (x_i - \overline{x})
  (y_i - \overline{y})}
  { \{ \sum_{i = 1}^{M} 
  (x_i - \overline{x})^{2} \}^{1/2} \, 
  \{ \sum_{i = 1}^{M} 
(y_i - \overline{y})^{2} \}^{1/2} } .
\end{equation}

Contrary to an expectation from the standard picture that
a small binding energy leads to a large nuclear radius
due to the extension of the wave function in the asymptotic region,
the obtained correlation between $S_{2n}$ and $R_{\rm rms}$
becomes not so strong, i.e. $r_{xy} = -0.801$.
Therefore, a question arises,
``which physical quantity is the explanatory
variable to determine the nuclear radius?''

To answer this question,
we plot the calculated results of the 12 parameter sets
in Fig.~\ref{fig:E30F_Rrms} under the following manner.
We take the $x$-axis for
the energies of the $p$-orbit and the $f$-orbit in $^{30}$F
and the $y$-axis for the matter radius of $^{31}$F.
The symbols are assigned to different diffuseness parameters as
$a = 0.65, 0.70$, and $0.75$ fm,
and the distance between the solid and open symbols
connected by a dashed-line 
corresponds to the energy gap between the $p_{3/2}$ orbit
and the $f_{7/2}$ orbit in $^{30}$F.
As seen from  Fig.~\ref{fig:E30F_Rrms},
it can be considered that
the radius of $^{31}$F is strongly correlated to 
the energy gap in $^{30}$F. 

In order to confirm the above investigation,
we calculate PCC between the energy gap $\Delta\varepsilon$ in $^{30}$F 
and the radius $R_{\rm rms}$ of $^{31}$F.
Here, the energy gap is defined as
$\Delta \varepsilon \equiv \varepsilon(f) -  \varepsilon(p)$,
where $\varepsilon(f)$ and $\varepsilon(p)$ are the real part of
the resonant energies of the lowest orbits of $7/2^{-}$ and $3/2^{-}$
in $^{30}$F, respectively.
The obtained PCC shows very strong correlation as $r_{xy} = 0.935$, 
which is stronger than that obtained
for $S_{2n}$ and $R_{\rm rms}$ as $r_{xy} = -0.801$.
Therefore, we consider the energy gap  $\Delta\varepsilon$
is a key variable to indicate the structural change of $^{31}$F.

\begin{figure*}
  \includegraphics[width=0.73\textwidth]{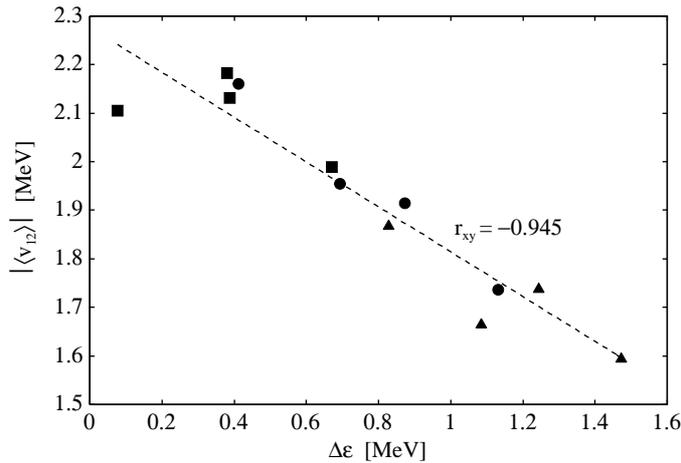}
  \caption{The correlation between the absolute value of
    the neutron-neutron interaction $\braket{\hat{v}_{12}}$
    in the $^{31}$F system
    and the energy gap $\Delta \varepsilon$ in $^{30}$F.
    The definition of the triangles, circles and squares
    are the same with those applied in Fig.~1.}
\label{fig:DeltaE_V12}    
\end{figure*}

The energy gap also has a close relationship 
to the expectation value of the neutron numbers
for the valence orbits in $^{31}$F.
We calculate these numbers for the $p_{3/2}$ orbit ($N(p)$)
and the $f_{7/2}$ orbit ($N(f)$), 
where $N(f)$ and $N(p)$ are added up to 2 including other orbits.
In the largest radius case $R_{\rm rms} = 3.70$ fm,
$N(p)$ and $N(f)$ are obtained as $1.63$ and $0.19$, respectively,
i.e. $^{31}$F is $p$-wave dominant.
On the other hand, in the smallest radius case  $R_{\rm rms} = 3.48$ fm, 
$^{31}$F becomes $f$-wave dominant as $N(p) = 0.38$ and $N(f) = 1.53$.
As shown in Fig.~3 of Ref.~\cite{Ma20_PRC101},
$N(p)$ and $N(f)$ change inseparably and intersect each other 
at about $\Delta\varepsilon = 0.4$ MeV.
Hence, it is confirmed that 
the energy gap $\Delta\varepsilon$ is related to the neutron numbers
and also to the radius.
However, even in the smallest radius case,
a small separation energy $S_{2n} = 1.37$ MeV
leads to the small radius $R_{\rm rms} = 3.48$ fm.
Therefore, we consider a new mechanism to shrink
the nuclear radius should occur in $^{31}$F,
which can be an ``anti-halo'' effect.

Next, we investigate how the energy gap affects
to the pairing strength in the Borromean system,
since the anti-halo effect has been discussed in terms of
the pairing correlation~\cite{Be00_PLB496,Ha11_PRC84}.
Figure~\ref{fig:DeltaE_V12} shows the absolute value
of $\braket{\hat{v}_{12}}$ in the Hamiltonian (\ref{eq:COMS_Hamiltonian})
with respect to the change of the energy gap $\Delta\epsilon$.
As shown in Fig.~\ref{fig:DeltaE_V12},
$\Delta\epsilon$ is strongly correlated to $| \braket{\hat{v}_{12}}|$,
where PCC is obtained as $r_{xy} = -0.945$.
It is noted that the correlation
between the separation energy  $S_{2n}$ and  $|\braket{\hat{v}_{12}}|$
is very weak by considering PCC obtained as $r_{xy} = 0.660$.
Combined with the result for
the $\Delta\varepsilon$ dependence of $N(p)$ and $N(f)$,
we can confirm that 
$|\braket{\hat{v}_{12}}|$, which can be an index of
the neutron-neutron correlation,
increases for a small energy gap,
and $^{31}$F becomes $f$-wave dominant.
As a result, the radius of $^{31}$F shrinks
even though the separation energy is still small.

To realize the strong correlations between $\Delta\epsilon$
and other quantities such as $R_{\rm rms}$, $N(p)$, $N(f)$,
and $|\braket{\hat{v}_{12}}|$,
and the shrinkage of the radius induced by $\Delta\epsilon$,
there is another important condition
that is the inversion of the valence orbits.
$^{30}$F and $^{31}$F are considered to be placed on the ``island of inversion'',
where the order of the single-particle orbits is inverted
from the normal shell-model one.
If the $f$-orbit ($f_{7/2}$) lies above the $p$-orbit ($p_{3/2}$), 
a competition occurs
between the energy loss from the gap $\Delta\varepsilon$
and the gain from the nucleon-nucleon
interaction $|\braket{\hat{v}_{12}}|$,
and consequently, $\Delta\varepsilon$ becomes an explanatory variable to determine
$|\braket{\hat{v}_{12}}|$, $N(f)$, $N(p)$, and $R_{\rm rms}$.

\begin{figure*}
  \includegraphics[width=0.75\textwidth]{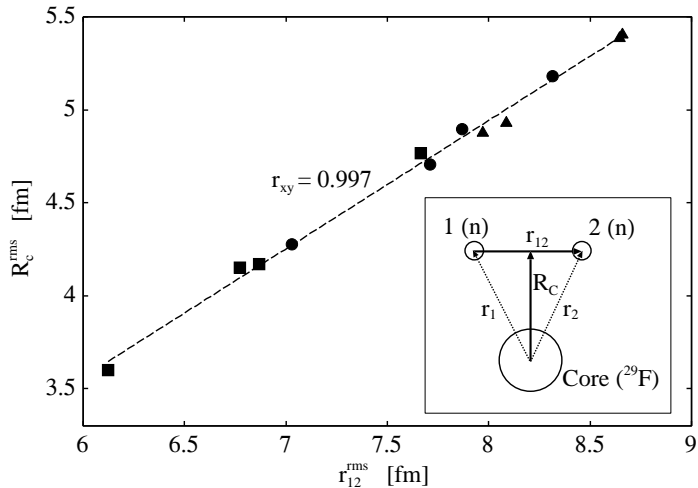}
  \caption{The correlation between
    $r_{12}^{\rm rms} \equiv \sqrt{\braket{\bv{r}_{12}^{2}} }$
    and $R_{c}^{\rm rms} \equiv \sqrt{\braket{\bv{R}_{c}^{2}}}$.
    The right-bottom panel shows the coordinate system
    of the $^{29}$F+$n$+$n$ three-body system
    in COSM ($\bv{r}_{1}$ and $\bv{r}_{2}$), 
    and the definition of $\bv{r}_{12}$ and $\bv{R}_{c}$.
    The definition of the triangles, circles and squares
    are the same with those applied in Fig.~1.}
\label{fig:r12_Rc}    
\end{figure*}

From the above discussion, the key ingredient that determines
the nuclear radius of $^{31}$F is considered to be the energy gap
between the $p_{3/2}$ orbit and the $f_{7/2}$ orbit in $^{30}$F.
The formation of the halo structure in $^{31}$F
depends on whether the energy gap is large enough
to overcome the energy loss of the nucleon-nucleon interaction part.
Such the situation generally occurs for systems
that the valence particle orbitals are inverted,
i.e. a smaller angular momentum orbit lies
below a higher angular momentum orbit.
As the energy gap decreases,
neutrons begin to occupy the orbit with the higher angular momentum,
and the nucleon-nucleon interaction part gives larger contribution
to overcome the energy loss from the gap.
As a result, the radius becomes small even for a small separation energy.
We consider this mechanism is a novel anti-halo effect.

Finally,
in addition to the anti-halo effect,
we investigate the possibility of the di-neutron like localization in $^{31}$F.
The correlation between the neutron-neutron distance
$r_{12}^{\rm rms} \equiv  \sqrt{\braket{\bv{r}_{12}^{2}}}$
and distance from the core to the center of mass of valence
neutrons $R_{C}^{\rm rms} \equiv \sqrt{\braket{\bv{R}_{c}^{2}}}$
is shown in Fig.~\ref{fig:r12_Rc}.
If the ratio of  $r_{12}^{\rm rms}$ to $R_{c}^{\rm rms}$
deviates from the line, for example, in the case of 
a small  $r_{12}^{\rm rms}$ and a large $R_{c}^{\rm rms}$,
the neutrons are supposed to have a di-neutron like localization.
However, the correlation becomes very strong as
$r_{xy} = 0.997$, which is almost the liner correlation,
and it is difficult to find a possibility of the di-neutron like localization
for the parameter sets of this calculation.


\section{Summary}
\label{sec:4}
We study the structure of $^{31}$F,
which is the drip-line nucleus of the fluorine isotopes,
using COSM with GEM in a $^{29}$F+$n$+$n$ three-body model.
Since there is an ambiguity to determine the $^{29}$F+$n$ potential,
the strength parameter of WS can be changed
by keeping the consistency for $^{31}$Ne and $^{30}$F.
The correlation analysis is performed from the results
using the 12 different parameter sets by keeping
the consistency.
As a result, we found that the key ingredient
to determine the nuclear radius of $^{31}$F
is the energy gap between the $p_{3/2}$ orbit and $f_{7/2}$ orbit in $^{30}$F.
This is considered as a general condition for other nuclei near
the drip-line, where the valence orbits are inverted,
and it is expected that the cross over of the halo structure
and the occurrence of the anti-halo effect can be determined 
by the energy gap between these inverted valence orbits.

\begin{acknowledgements}
This work was in part supported by JSPS KAKENHI
Grant Nos. 18K03636, 18K03635, 18H04569, 19H05140, and
19K03859, and the collaborative research program 2021,
information initiative center, Hokkaido University.
\end{acknowledgements}


\end{document}